\documentclass[a4paper,10pt,preprint,3p,twocolumn,number,sort&compress,times]{elsarticle}
\usepackage[utf8]{inputenc}
\usepackage{graphicx}
\usepackage{xcolor}
\usepackage{subcaption}
\usepackage{caption}
\usepackage{soul}
\usepackage{lineno}
\PassOptionsToPackage{hyphens}{url}\usepackage[breaklinks]{hyperref}
\usepackage{breakurl}
\journal{Journal of Alloys and Compounds}

\address[MFFUK]{Charles University, Faculty of Mathematics and Physics, Department of Condensed Matter Physics, Ke Karlovu 5 121 16 Praha 2, Czech Republic}
\address[FZU]{Institute of Physics, Academy of Science of the Czech Republic, Cukrovarnická 10, 162 00 Praha 6, Czech Republic}

\begin{document}
\begin{frontmatter}

\title{Lattice distortion in TmCo$_2$: a poly- and single- crystal study}

\author[MFFUK]{J.~\v{S}ebesta\corref{cor1}}
\cortext[cor1]{Corresponding author}
\ead{sebesta.j@email.cz}

\author[MFFUK,FZU]{D.~Kriegner}

\author[MFFUK]{J.~Prchal}

\begin{abstract}

Within the RCo$_2$ family of compounds, a structural distortion linked with the onset of magnetic ordering around the critical temperature can be observed. One of the less explored RCo$_2$ compounds is TmCo$_2$ probably due to its low Curie temperature. 
%
%%%%
Exceptionally this compound, given its position at the end of the ferrimagnetic series, shows discrepancies in the ordering of the Co sub-lattice because of a weak Weiss molecular field. In this paper we focus on the structural distortion in TmCo$_2$, which appears together with the magnetic ordering around the critical temperature of $T_{C}\sim3.6$~K.
Poly-crystals as well as single-crystals of TmCo$_2$ were used in our experiments. For both kinds of samples we observed the same type of the rhombohedral distortion along the [111] direction from the cubic Fd$\bar{3}$m to R$\bar{3}$m space group. The relation between observed magnetic and structural properties in this compound is discussed.

\end{abstract}

\begin{keyword}
   magnetically ordered materials \sep rare earth alloys and compounds \sep magneto-volume effects \sep X-ray diffraction
 \PACS  61.05.cp \sep 61.50.Ks \sep 75.50.Gg \sep 75.30.-m 
\end{keyword}

\end{frontmatter}

%\linenumbers

\section{Introduction}
The RCo$_2$ compounds (where R represents a rare-earth element) crystallize in the Fd$\bar{3}$m (number 227) type of space group with prototype MgCu$_2$, which belongs to the group of so-called Laves phases. The Laves phases are compounds with a closed packed structure, which leads to specific physical properties (see e.g. \cite{Anton}).
In this structure, R atoms occupy the positions of a diamond lattice. Between them the Co atoms form regular tetrahedra. From the magnetic point of view the RCo$_2$ compounds are interesting due to a coexistence of localized magnetic moments on the R-element site as well as itinerant magnetic moment on the Co site. While the R atoms have magnetic moments due to the partly filled $4f$-electron shells, Co moments arise from the splitting of the $3d$-electron bands. Moments having origin in these bands appear on the verge of magnetism in the sense of the Stoner criterion \cite{AshcroftM}. % 
The R-sublattice produces a strong Weiss molecular field, strong enough to split the Co-bands. This makes this family of compounds very sensitive to external conditions, which may shift the cobalt subsystem to the magnetic state \cite{Goto}. The formation of a cobalt itinerant magnetism is accompanied by a first order magnetic phase transition (FOMPT), which is not common for magnetic phase transitions, because they are usually of the second order \cite{Goto}. The existence of a FOMTP in RCo$_2$ compounds is connected with the presence of the magnetocaloric effect, which is one of reasons, why these compounds were studied in the past \cite{Wada}.

Depending on the rare-earth element R there are observed two ferromagnetic sub-lattices below the Curie temperature $T_C$ - one formed by the rare-earth atoms and the other one by the cobalt atoms. These sublattices order parallel with respect to each other for the light rare-earths (R=Pr, Nd, Sm), however, for the heavy ones (R=Gd, Tb, Dy, Ho, Er) anti-parallel, forming a ferrimagnetic 
structure \cite{Driver}. In case of non-magnetic Y, Lu elements one can find Pauli paramagnets with a metamagnetic transition \cite{Driver,Gignoux}.

Besides the ferro- or ferrimagnetism the RCo$_2$ compounds exhibit also ferroelastic properties. The magnetic ordering below $T_C$ and the changes in the magnetic ordering have an influence on the crystal structure of the compound leading to a spontaneous breaking of the symmetry \cite{Driver}. Together with the magnetic ordering a lattice relaxation leading to the transformation of the crystal structure takes place. Here the type of the transformation again depends on the type of the R-element. 
In the temperature dependence of the easy magnetization axis in the magnetically ordered phase one can observe few types of the structural distortion. ErCo$_2$ and NdCo$_2$ could be chosen as the representative well studied compounds. 
%%%%%%%%  
{ In the ordered state of ErCo$_2$ the easy axis evolves along one of the equivalent [111] direction, which changes the structure to the rhombohedral one with the crystallographic space group R$\bar{3}$m and easy axis along [111] direction \cite{Kozlenko}.} Here the mentioned crystallographic directions refer to the initial cubic structure. We will use this notation further in the text if not explicitly stated otherwise.
The case of NdCo$_2$ is more complicated due to the existence of an easy axis reorientation at the temperature $T_R$ below $T_C$ \cite{Xiao,Ouyang,Gratz83}. Here the structure changes from the initial cubic lattice via an orthorhombic lattice with the [110] easy axis direction for $T_R \le T \leq T_C$ to a tetragonal lattice connected with the easy axis in the [100] direction for $T \leq T_R$. {Related }the {respective} space group changes accordingly from Fd$\bar{3}$m to I4$_1$adm {at $T_C$} and further to Imma {below $T_R$}. 

One of the more complicated examples among the RCo$_2$ compounds is TmCo$_2$, probably due to the low ordering temperature and also technical issues due to high evaporation of Tm during  preparation\cite{Gratz95,Baranov}. TmCo$_2$ is situated at the end of the series of ferrimagnetic RCo$_2$ compounds \cite{Driver}. This is connected with the vicinity of the critical molecular magnetic field to establish the cobalt magnetism  \cite{Gratz95,Gratz96,GratzMarkosyan} as it was shown in Ref.~\cite{Goto}. In this reference the critical molecular field is related to
the metamagnetic transition in the analogue YCo$_2$. This fact leads to discrepancies in the literature, where one can find results with a magnetic \cite{Nakama,Gignoux,Deportes,Bonilla}  as well as non-magnetic cobalt {sublattice} \cite{Gratz95,Gratz96,Baranov,GratzMarkosyan,Dubenko,Resel}.
These two cases are ascribed to the observation of a FOMPT \cite{Gubbens} or SOMPT (second order magnetic phase transition) respectively. Also the published results differ with respect to the value of the Curie temperature, which varies from 3.8~K to roughly 7~K. Moreover some articles inconsistently 
show additional anomalies present close to $T_C$. 
To solve {these} obvious existing discrepancies and to gain deeper insight into the behavior of TmCo$_2$ we have prepared this compound several times in both – poly-crystalline and single-crystalline form and performed low-temperature magnetic and structural investigations. Our results shown below demonstrate that the rhombohedral distortion at the ordering temperature is present independently on the type of the transition and that the first-order type of transition has its origin in the Co metamagnetic transition.

\section{Experimental Details}

\subsection{Sample Preparation}

Annealed and well characterized poly-crystalline samples were
used in the experiments. Preparation and characterization of these samples was reported in Ref. \cite{Sebesta}. {In} addition a single-crystalline sample was studied. {It was successfully prepared after several attempts by the Bridgman method.} For the single-crystal preparation the initial temperature was set to 1410$^\circ$C, which was obtained from differential scanning calorimetry (DSC). 
%%% 
After multiple attempts we determined the suitable pulling rate to be 3\,mm per hour and the sample was annealed at 900$^\circ$C for 10 days after the preparation, equal to the procedure of the poly-crystal.

\subsection{Characterization}
The prepared single-crystalline sample was characterized by a microprobe, which corroborates the presence of the TmCo$_2$ phase with only a subtle amount of Tm-rich impurities. A {Laue} X-ray diffraction (XRD) pattern was also measured giving the suitable result for this compound. 
From the obtained XRD patterns the lattice parameter $a=7.137(3)$~Å was determined in reasonable agreement with the poly-crystalline sample ($a_{poly}=7.142(2)$~Å) and the values reported in literature \cite{Liu_Altounian}. 

The bulk physical properties – magnetization, electrical resistivity, and thermal expansion – were measured by standard methods using a commercial PPMS (Quantum Design). A common four probe method was used for the measurement of the resistivity. For thermal expansion measurement a microdilatometric capacitance cell implemented to the PPMS was used \cite{Rotter}. The oriented plan-parallel samples were placed to the core of the cell.
Further we performed low-temperature XRD (LT-XRD) experiments. For that purpose we used the Bragg-Brentano $\Theta$-2$\Theta$ geometry in a refurbished Siemens D500 diffractometer. Cooling of the sample was provided by a closed-cycle refrigerator (CCR, Sumitomo Heavy Industries), and He exchange gas ensured {equalization} of the temperature between the cold-finger, thermometer and sample. $Cu-K_{\alpha1,2}$ radiation and a linear detector was used to speed up the data recording \cite{Kriegner}. For the poly-crystalline samples the obtained diffraction patterns were analyzed using the Rietveld method implemented in the FullProf program \cite{Carvajal}. The single-crystalline sample was measured with the same setup but using an additional sample rotator implemented inside the cryostat (Attocube) to orient the single-crystal to the appropriate measurement geometry. 

\begin{figure}
	\centering
	\includegraphics[width=0.5\textwidth]{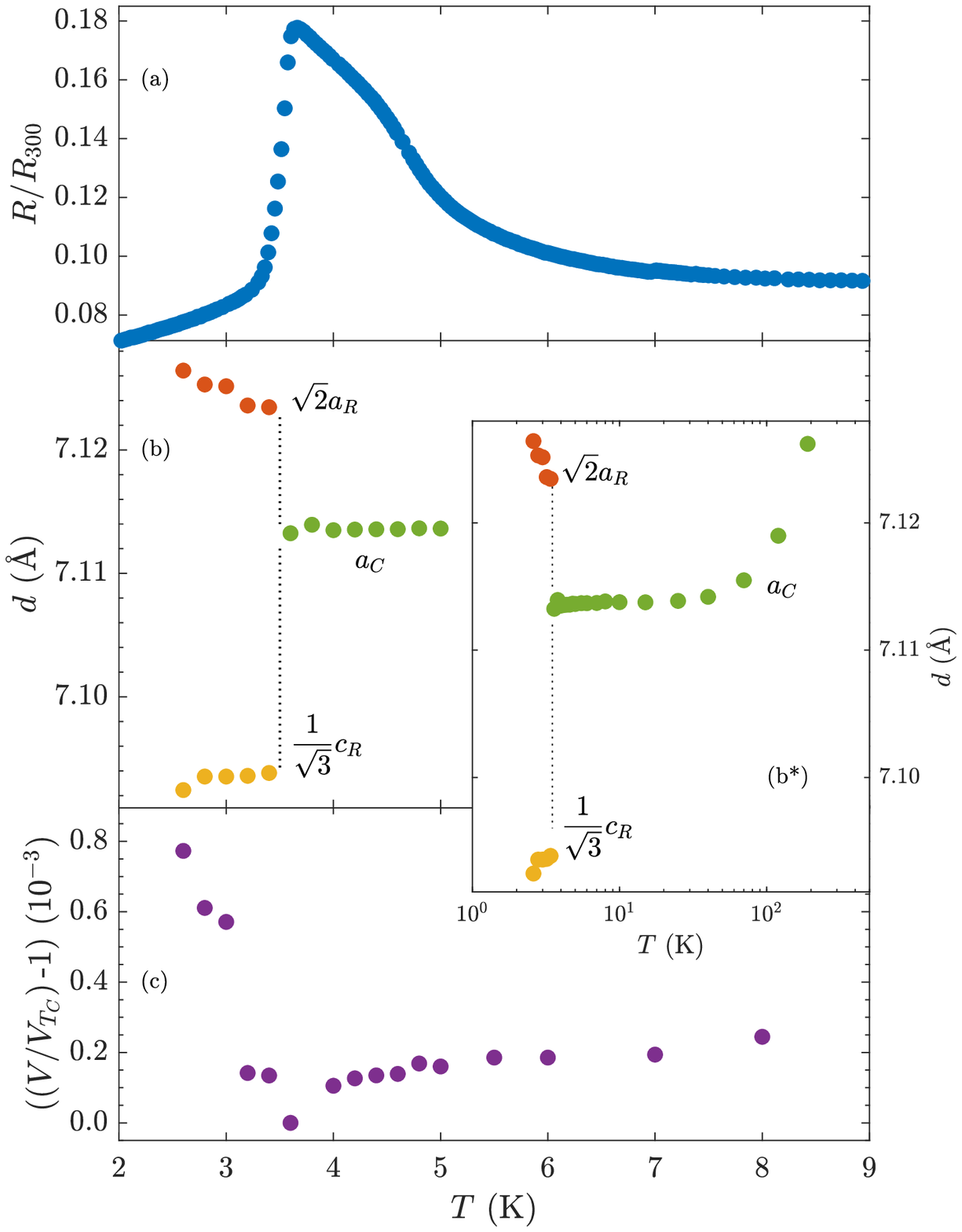}
	\caption{TmCo$_2$ poly-crystal data:   \newline 
a) temperature dependence of the electric resistivity, 
b) temperature dependence of lattice parameters; $a_C$ cubic lattice parameter; $a_R$,~$c_R$~rhombohedral lattice parameters 
c) evolution of the relative cell volume}
    \label{fig:1}

\vspace{30pt}

	\centering
	\includegraphics[width=0.47\textwidth]{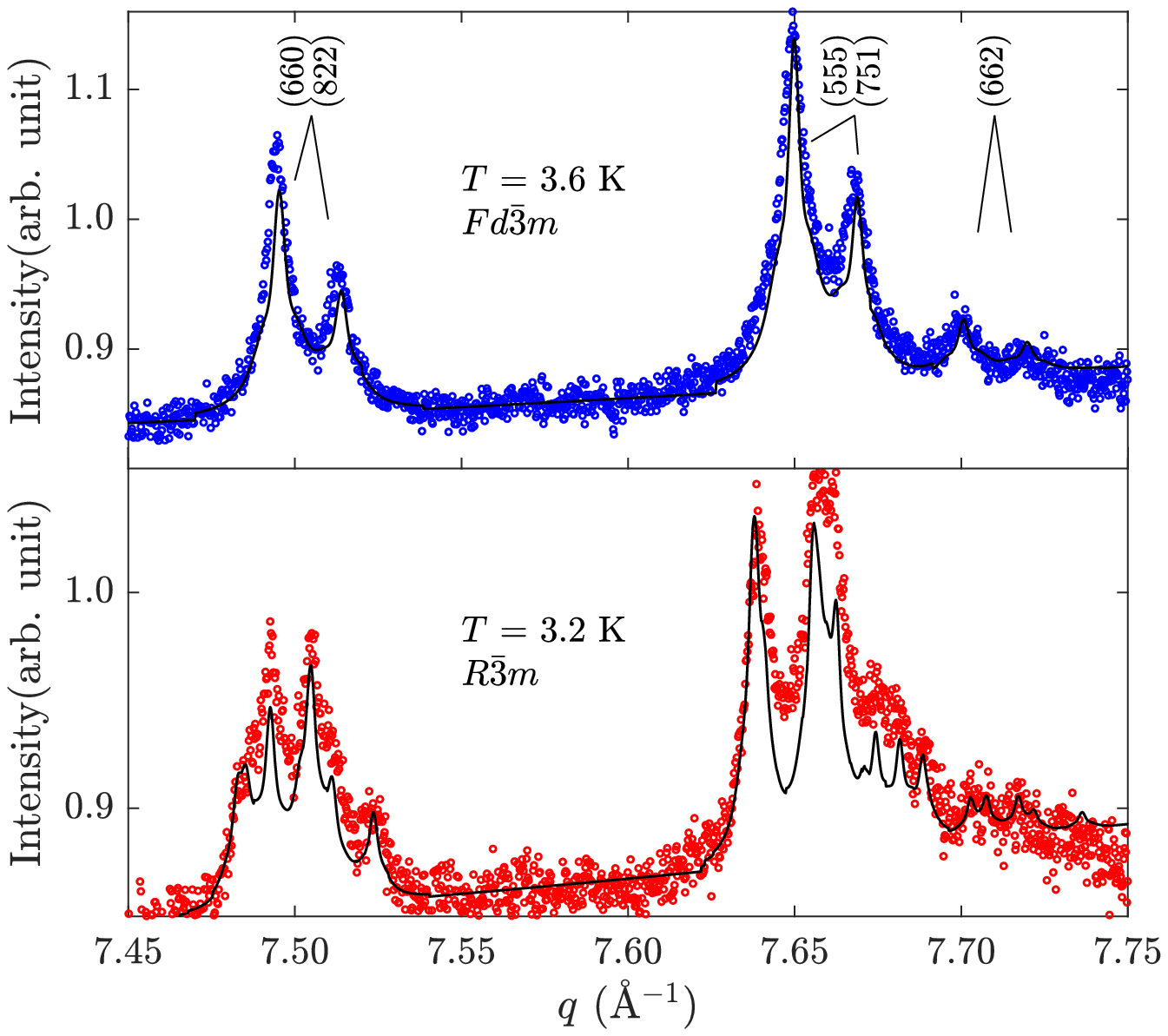}
	\caption{Detail of the XRD patterns of the TmCo$_2$ poly-crystal above and below the magnetic ordering temperature.}
    \label{fig:2}

\end{figure}

\section{Results}

\subsection{Poly-crystal}

On the poly-crystalline sample low temperature X-ray powder diffraction was measured {down to} the lowest possible temperature around 2.6~K, i.e. below the assumed Curie temperature. The lattice parameter at room temperature was fitted as 7.137~Å and it decreases with decreasing temperature to 7.113~Å at the ordering temperature $T_{C}$ (as shown in the (Fig.~\ref{fig:1}b), while the sample preserves the cubic structure with a crystallographic space group Fd$\bar{3}$m. Further obtained results evidence a lattice distortion and changing of the symmetry in the vicinity of the Curie temperature $T_{C}$.
When the temperature drops below $T_C$ the measured pattern exhibits a splitting of cubic reflections (e.g. $(555)$, $(751)$) (Fig.~\ref{fig:2}). According to the profiles of the reflections and their shifting we find that these changes are caused by the rhombohedral distortion (Fig.~\ref{fig:2}).

The crystal structure changes from the cubic structure with the space
group Fd$\bar{3}$m due to the rhombohedral distortion to the trigonal
structure with R$\bar{3}$m space group. The rhombohedral distortion causes the shortening along the [111] direction. In the following we will use the hexagonal notation to describe the trigonal system for simplicity.
The structural transition is apparent from the comparison of the cubic and hexagonal lattice parameters (Fig.~\ref{fig:1}b). The change of structure is denoted there by the substantial steps deviating from the evolution caused by thermal expansion. The displayed parameters in (Figs. \ref{fig:1}b,~\ref{fig:1}b$^\ast$) are projections of the trigonal system to the parent cubic one.

\subsection{Single-crystal}

\begin{figure} [t]
	\centering
	\includegraphics[width=0.45\textwidth]{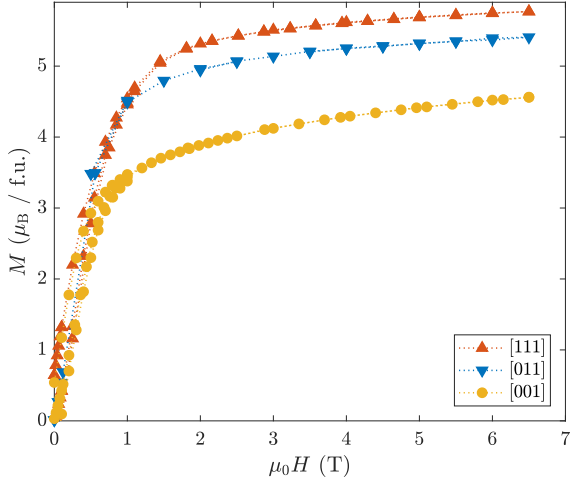}
	\caption{Magnetization curves of the TmCo$_2$ single-crystal along the principal directions at $T=2$~K}
	\label{fig:3}
\end{figure}

Magnetization curves were measured on the single-crystalline samples in the prominent crystallographic directions to find out the magnetic easy axis. The measurements showed that at the lowest temperature of 2\,K the easy axis of the magnetization lies in the [111] direction and the hard one in the [001] direction (Fig.~\ref{fig:3}). This finding is in agreement with previously reported \cite{Gignoux,Deportes}.

\begin{figure} [t]
	\centering
	\includegraphics[width=0.5\textwidth]{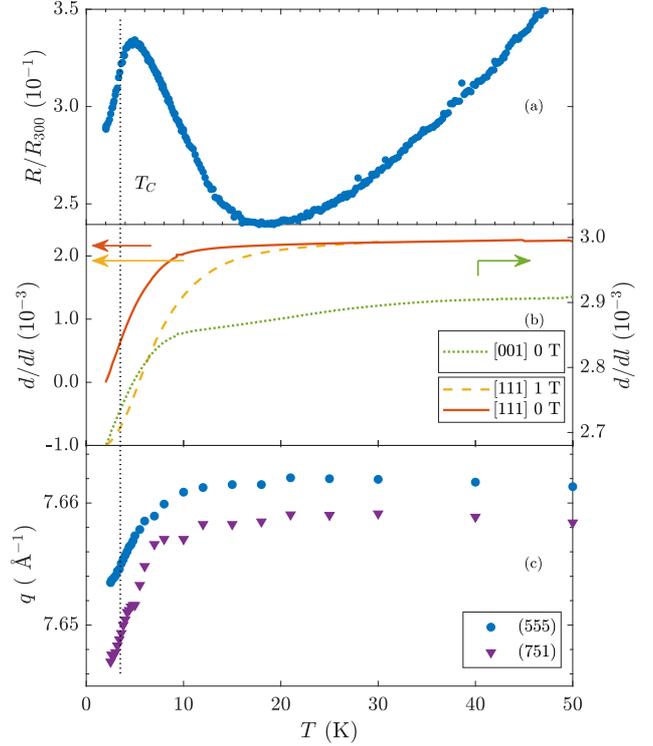}
	\caption{TmCo$_2$ single-crystal data:\newline 
a) temperature dependence of the electric resistivity
b) temperature dependence of the thermal expansion
c) fitted positions of the respective reflection from the LT-XRD measurement}
	\label{fig:4}
\end{figure}

In order to determine the ordering temperature of the single-crystalline sample we measured the temperature dependence of the electric resistivity (Fig.~\ref{fig:4}a) and dilatometry (Fig.~\ref{fig:4}b). An inflection point of the significant drop in the acquired temperature dependence of the electric resistivity was denoted as the ordering temperature. The derived value of the ordering temperature is 3.5(1)~K. The shape of the resistivity dependence indicates a second order phase transition (Fig.~\ref{fig:4}a).

In the case of the single-crystalline sample the dilatometry (Fig.~\ref{fig:4}) and the low temperature X-ray diffraction was measured. The thermal expansion was measured along the easy [111] direction and the hard [001] direction. The benefit of this method is high sensitivity, which can be better than XRD measurements \cite{Henriques_JMMM}. 
The obtained temperature dependences of the thermal expansion shown in Fig.~\ref{fig:4}{b} do not include such expressive transition as in the case of powder XRD data on the poly-crystalline sample (Fig.~\ref{fig:1}{b}). Nevertheless, they evidence a significant change of the measured directions, but extended over a broader temperature interval. 
%%% 
In the case of zero external field the data evidence the transformation, which begins around 10~K and the maximum of the thermal expansion coefficient $\alpha$ occurs at 3.8(1)\,K for both directions {(Fig.~\ref{fig:4}b)}.
It is necessary to point out that in the case of the hard direction the detected change was about one order of magnitude smaller than along the easy axis. The external magnetic field applied parallel to the magnetic easy axis direction shifts the transition to higher temperatures (Fig.~\ref{fig:4}b)

\begin{figure*}

\begin{minipage}{.495\textwidth}
	\centering
	\includegraphics[width=0.92\textwidth]{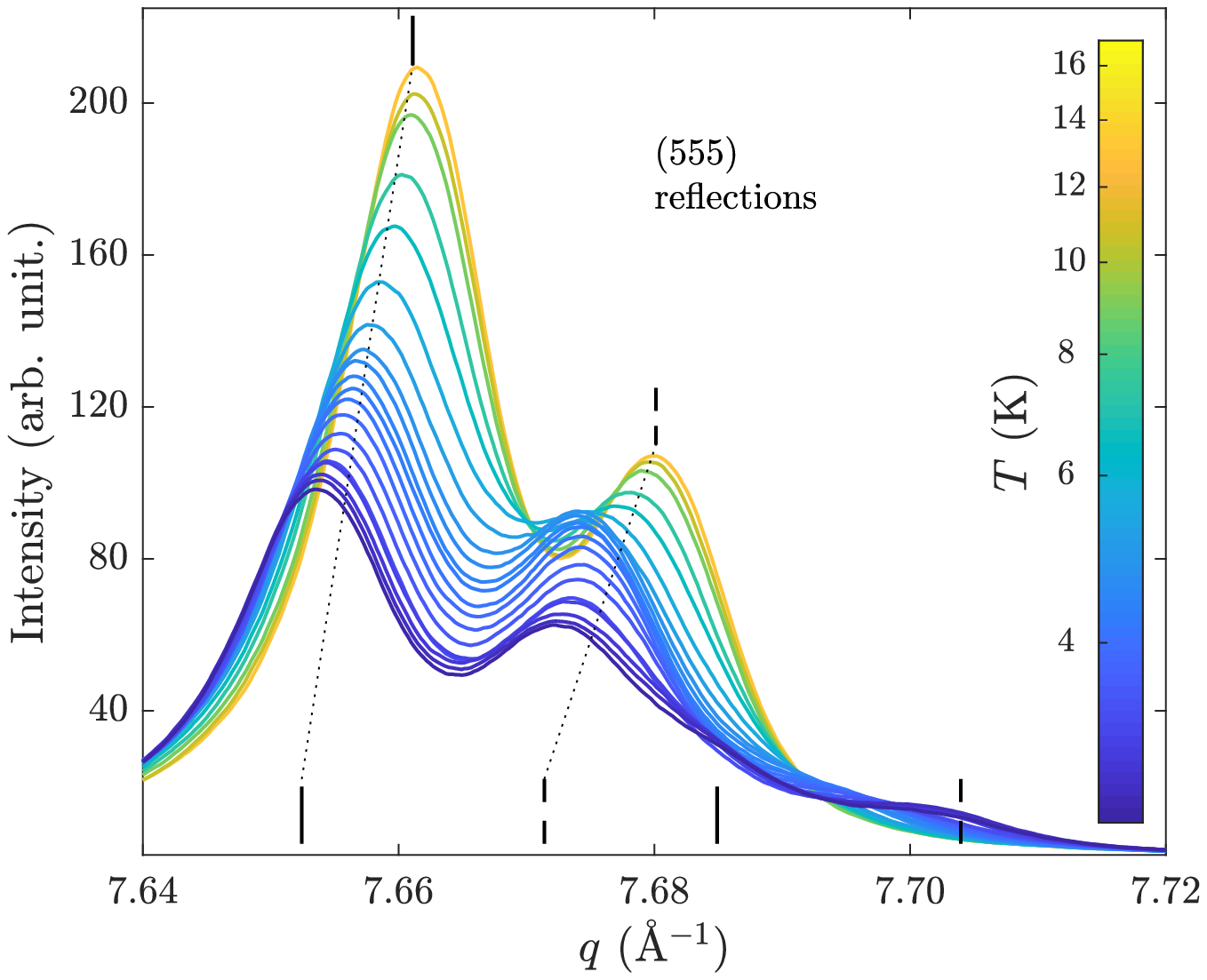}  
    \caption{TmCo$_2$ single-crystal: Temperature evolution of the intensity of (555) reflections integrated along the rocking angle.
Vertical lines denote the positions of corresponding diffraction peaks in the poly-crystal sample. Their origin is distinguished by a line style.}
	\label{fig:5}

\end{minipage}
\hfill
\begin{minipage}{.495\textwidth}

	\centering
	\includegraphics[width=0.93\textwidth]{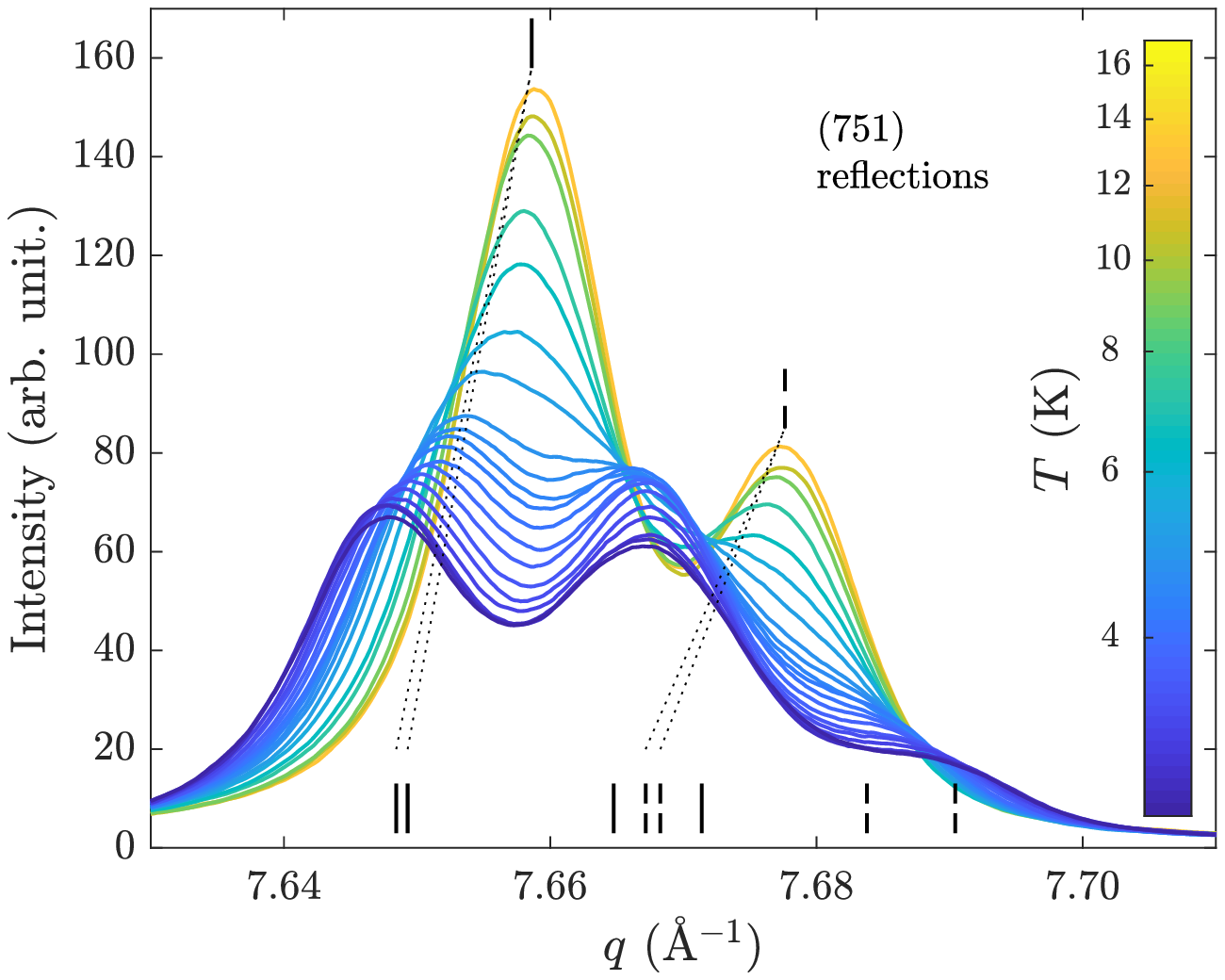}
	\caption{TmCo$_2$ single-crystal: Temperature evolution of the intensity of (751) reflections integrated along the rocking angle.
Vertical lines denote the positions of corresponding diffraction peaks in the poly-crystal sample. Their origin is distinguished by a line style.}
	\label{fig:6}

\end{minipage}
\end{figure*}

For a more detailed description of the observed structural transition the measurement of the low temperature diffraction on single-crystalline samples was performed. From the XRD pattern obtained on the poly-crystalline sample we selected the reflections suitable for the investigation of the single-crystalline sample. Therefore the reflections (555) and (751) close to the momentum transfer $q=4\pi/\lambda \sin\theta = 7.65$~Å$^{-1}$, with the wavelength of the used X-rays $\lambda$, and the Bragg angle $\theta$ were chosen. The selected Bragg reflections exhibit a noticeable splitting and additionally can apriori not be distinguished in the poly-crystalline powder x-ray diffraction measurements since they both contribute to the signal at the same Bragg peak position. Special attention was required to find the appropriate orientation of the single-crystal. The plan-parallel sample with the plane oriented to the [111] direction was inserted to the diffractometer. At each temperature point we measured $\omega/2\theta$ scans which together with the linear detector yield reciprocal space maps of the vicinity of the selected Bragg peak. To obtain diffraction patterns similar to the powder one, we integrate the measured intensities along the rocking angle. For both reflections, (555) and (751), the obtained patterns exhibit evident splitting below 10~K (Figs.~\ref{fig:5},~\ref{fig:6}).

We fitted the position of the dominant peaks with decreasing temperature to demonstrate a sudden change of their position in the vicinity of the critical temperature (Fig.~\ref{fig:4}c). We assume that this feature represents a continuous structural transition given by a magnetic ordering and the fitted peak positions clearly describe its evolution.

\begin{figure} [h]
	\centering
	\includegraphics[width=0.44\textwidth]{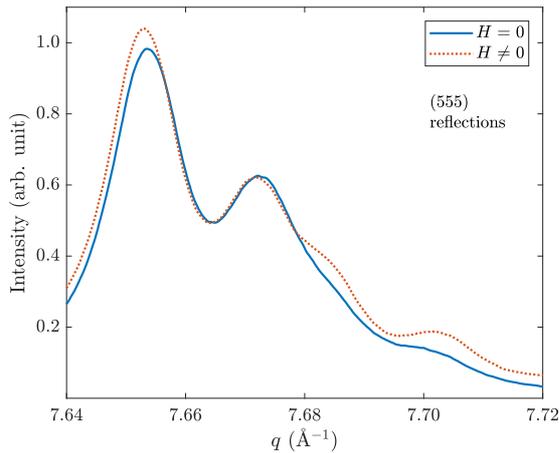}
	\caption{Comparison of the obtained diffraction patterns of the TmCo$_2$ single-crystal at the lowest measured temperature of 2.6~K as measured without external magnetic field and with the magnetic field $H$ applied along [111]. An influence of the external magnetic field to the XRD patterns measured below the magnetic ordering temperature is visible.}
	\label{fig:7}
\end{figure}

We made an attempt to increase the visibility of the structural change by inserting a small permanent magnet {under} the measured sample. The magnetic field was aligned parallel to the [111] direction. This direction is the magnetic easy axis and we expected influencing the population of the magnetic domains. In particular to increase the population of the domains oriented along the specific [111] direction parallel to the surface normal. Due to restricted sample space we were able to use only (Nd-Fe-B) magnet with about 1~mm thickness and 3~mm in diameter. Although the magnetic field was not very strong, the obtained XRD pattern (Fig.~\ref{fig:7}) evince more pronounced peaks, which are slightly shifted in comparison to the initial ones (i.e. without magnet) indicating probably the stronger distortion when a magnetic field is applied.

\section{Discussion}

The results of the LT-XRD on the poly-crystalline TmCo$_2$ (Fig.\ref{fig:1}) correspond {well} with the previous measurement of the electric resistivity, heat capacity and AC magnetic susceptibility performed on this sample \cite{Sebesta}.
%%%
The structural transition is strictly described by the sudden splitting of specific reflections in the XRD patterns (Fig.~\ref{fig:1}). The change of the structure from the cubic to the trigonal {one} (Fig.~\ref{fig:2}) corresponds to the results obtained on the related compounds in the literature \cite{Gratz94}. It is in agreement with the discovered magnetic easy axis, because the change of the structure type is related to the magnetic ordering and the easy axis direction affects the type of the distortion \cite{Driver,Gratz94,Gratz83}. Also the observed structural distortion confirms the existence of the first order phase transition, which was detected in the previous resistivity and heat capacity measurements \cite{Sebesta}. {T}he first order type of transition is eviden{t in our data} by the sudden change of the crystal structure (Fig.~\ref{fig:1}b).

The occurrence of the first order phase transition is crucial to the presence of the itinerant Co-magnetism in this matter. It should indicate the formation of the itinerant band meta-magnetism in the Co-sublattice. In literature, however, also an opposing opinion exists, which states that the origin of the FOMPT represents a gain of energy due to volume expansion related to the structural transition \cite{Gratz95}.
Based on a dilatometry experiment a negative volume change connected with the magnetic ordering was observed \cite{Gratz95}. This fact is in the contradiction to the appearance of the Co-magnetic moment \cite{Gratz95,Muraoka} and it leads to an insufficient magnitude of Weiss molecular field. Hence related compounds doped by another magnetic rare-earth element were studied, where the doping element increases the effective magnetic field above the critical value to form Co-magnetism \cite{Gratz96}. Due to that a behavior similar to other heavy RCo$_2$ appeared.

Contrary to that, we observed a positive volume change connected with the magnetic ordering in a pure TmCo$_2$ sample in our XRD experiment (Fig.\ref{fig:1}c). Its magnitude is close to the value presented for a Tm$_{1-x}$Gd$_x$Co$_2$ systems \cite{Gratz95}. We derived that the relative volume change is greater than 0.6$\cdot$10$^{-3}$. This indicates that the observed FOMPT in our case likely originates in the meta-magnetic transition in the Co-sublattice \cite{Gratz95}. %
%%% 

{In previous paragraphs it was shown that for the poly-crystal both sub-lattices are magnetic, which is reflected by FOMPT. The case of single-crystal seems to be quite different.} 
Although we get a similar ordering temperature for both types of samples, no such sharp transitions as for the poly-crystal {(Fig.~\ref{fig:1})} were observed for the single-crystalline sample (Fig.~\ref{fig:4}).  In the case of the temperature dependence of the electric resistivity the ordering temperature 3.5(1)~K was obtained, but the decrease in the resistivity was approximately ten times broader then in the case of the poly-rystalline sample. Similarly for the dilatometry data we get the inflection point close to 3.5~K, but the data exhibit a continuous change similar to the position of diffraction peaks from the low-temperature XRD (Fig.~\ref{fig:4}).  Moreover, the data from the experiments of dilatometry and XRD point to comparable temperature, where the transition may begin. It follows that the single-crystal doesn’t exhibit the first order phase transition, which is crucial for the existence of the magnetic Co-sublattice. This contradiction may be clarified by the discrepancies appearing in the literature \cite{Nakama,Gignoux,Deportes,Bonilla, Gratz95,Gratz96,Baranov,GratzMarkosyan,Dubenko,Resel}.

We have devoted a special effort to preparation of the samples with an emphasis on the sample quality and characterization. The quality of the prepared samples was found to be in a good state, especially in a comparison to the previously prepared sample of TmCo$_2$ or other RCo$_2$ samples showing very consistent results with the ones in literature. Nevertheless, we got the results showing differences between the TmCo$_2$ single-crystalline sample and the poly-crystal, which resembles the discrepancies found in the literature. We suppose that it is caused by the vicinity of the critical Weiss molecular field to establish the Co-magnetism \cite{Goto}, which implies probably a strong sensitivity of the matter to the certain composition (including local one) and the history of its preparation. This idea should also explain the discrepancies in the literature that these differences in the observed behavior of various TmCo$_2$ samples are consequences of the variations in the preparation.
Besides these discrepancies the studied samples have few aspects in common, i.e. show a similar ordering temperature, and the same type of structural distortion induced by the magnetic order.

To describe the character of the observed structural transition we tried to compare reflections of the poly-crystalline sample with the data obtained on the single-crystal (Figs.~\ref{fig:5},~\ref{fig:6}). For this it is necessary to differentiate, which reflections in the low temperature phase are related to the (555) reflections and which ones to the (751) reflections. In both cases the projections of the poly-crystalline reflections before and after the distortion agree quite well with the evolution of reflection observed in the single-crystal data. It seems that peaks in diffraction patterns (Figs.~\ref{fig:5},~\ref{fig:6}) follow split diffractions obtained for poly-crystal sample and which tends to appear at the lower $q$ values.

This means that the single-crystal also exhibits the rhombohedral distortion observed in data of poly-crystal with quite similar change of the lattice parameters. In addition, we observe one of the originally equivalent [111] directions, which is elongated during the transition, while another [111]-type direction is being contracted. It agrees with the distortion found in the poly-crystal (Fig.~\ref{fig:1}b). 
One should be aware that during the transition due to the symmetry change the originally equivalent [111]-type directions in the cubic system are not equivalent after the transformation any more.
This is supported by the significant contraction in the [111] direction observed in the dilatometry experiment in the same temperature region (Fig.~\ref{fig:4}b), which agrees with a reduction of hexagonal $c$-axis of distorted trigonal structure in the case of poly-crystal sample (Fig.~\ref{fig:1}b). These data are not inconsistent because the spontaneous breaking of the symmetry given during magnetic ordering depends on several conditions e.g. the shape anisotropy.
{The observed direction of the structural distortion is in agreement with the determined magnetic easy axis (Fig.~\ref{fig:3}). Therefore the symmetry breaking by the magnetic order is of the same nature as the structural distortion.}

Beside of the mentioned agreement we should stress that the similarity between diffraction results obtained for single-crystal and poly-crystal at the temperature of 2.6~K do not imply exactly the same structural transition. It would be contrary to the observed second-order type of the phase transition. The character of the single-crystal XRD data do not allow us to determine positions of reflections precisely.
%%%%%%%%%%%%%%%%
Nevertheless, there is no Co-magnetism in the case of single-crystal sample as the transition is of the second order type. We have observed  similar type of the structural transition as for the poly-crystalline sample, at which the transition is of the first order. This evidence indicates that the structural transition is primarily controlled by the rare-earth sublattice \cite{Levitin, Gratz94}.

The relation of the structure changes with the magnetic ordering is clearly demonstrated by the XRD experiment with the external magnetic field oriented along the [111] direction. There was observed a slight shift of the reflections arisen from a splitting of the cubic (555) reflections, which should represent an additional elongation of the hexagonal $c$-axis (Fig.~\ref{fig:7}). One can notice that the peak at lowest $q$ is slighly shifted to lower $q$ values. Also the other observed diffractions became better pronounced with application of magnetic field signifying increased domain-population with respect to the field-direction.

The direction of a magnetic field should promote ordering in the magnetic easy axis and influence the structural transition. The sensitivity of the XRD pattern to the external magnetic field proves the magnetoelastic behavior visible also in the dilatometry data. Although clearly visible, the magnitude of the change induced by the magnetic field is rather limited, which we explain by the fact that only a relatively small permanent magnet could be used.

\section{Conclusions}

The low temperature XRD experiment performed on TmCo$_2$ showed the existence of the rhombohedral distortion, which consists in the change from the  high temperature cubic phase described by the crystallographic space group Fd$\bar{3}$m  to the rhombic R$\bar{3}$m group. Our results in the case of the poly-crystalline sample confirm the previous results like an ordering temperature and  the occurrence of the FOMPT. It also indicates that the FOMPT has the origin in the Co meta-magnetic transition. This evidence is crucial and it shows that  the Co itinerant meta-magnetism exists also in the undoped TmCo$_2$ compound. Despite the single-crystalline sample exhibiting a continuous transition in all our measurements, which deny the existence of Co-magnetic sublattice, the result likely indicates the presence of the similar type of distortion. Observed differences between the samples illustrate the vicinity of the critical magnitude of the Weiss molecular field necessary to establish the Co-magnetism.

\section*{Acknowledgement}
The experiments were performed in MGML (http://mgml.eu/), which is supported within the program of Czech Research infrastructures (Project No. LM2018096).
The work was supported by the Grant Agency of Charles University, project GA UK No. 546214.
D.K. thanks to the project NanoCent - Nanomaterials centre for advanced applications
(Project No. CZ.02.1.01/0.0/0.0/15\_003/0000485) financed by the ERDF as well as the Ministry of Education of the Czech Republic Grant No. LM2018110 and LNSM-
LNSpin, the Czech Science Foundation Grant No. 19-28375X, and EU FET Open RIA Grant No. 766566.

\section*{References}

 \bibliography{reference}
 \bibliographystyle{elsarticle-num}

\end{document}